\documentclass[a4paper,fleqn]{cas-dc}

\usepackage[authoryear,longnamesfirst]{natbib}
\usepackage{mdframed}
\usepackage{listings}
\usepackage{graphicx} 
\usepackage{tcolorbox}
\usepackage{booktabs}
\usepackage{enumitem}
\usepackage{xspace}
\usepackage{xcolor}
\usepackage{makecell}
\usepackage{tabularray}
\usepackage[colorinlistoftodos]{todonotes}

\newcommand{\mypar}[1]{\noindent\textbf{#1}}
\newcommand{\mysubpar}[1]{\noindent\underline{\textit{#1}}}

\newcommand{\dsone}{\ensuremath{\mathcal{D}_1\,}}
\newcommand{\dstwo}{\ensuremath{\mathcal{D}_2\,}}
\newcommand{\malconv}{MalConv\xspace}
\newcommand{\gbdt}{GBDT-EMBER\xspace}
\newcommand{\pipeline}{SLIFER\xspace}

\lstdefinestyle{monospace}{
    basicstyle=\ttfamily,
    breaklines=true,
    frame=single,
}

\begin{document}
\let\WriteBookmarks\relax
\def\floatpagepagefraction{1}
\def\textpagefraction{.001}

\shorttitle{\pipeline: Investigating Performance and Robustness of Malware Detection Pipelines}

\shortauthors{Andrea Ponte et~al.}

\title [mode = title]{\pipeline: Investigating Performance and Robustness of Malware Detection Pipelines}                      

\author[1]{Andrea Ponte}[type=editor,
                        auid=000,bioid=1]

\cormark[1]

\ead{andrea.ponte@edu.unige.it}

\credit{Conceptualization of this study, Methodology, Software}

\affiliation[1]{organization={Department of Informatics, Bioengineering, Robotics and Systems Engineering, University of Genova},
    addressline={Viale Causa 13}, 
    city={Genova},
    postcode={16145}, 
    country={Italy}}

\author[1,2]{Dmitrijs Trizna}[]
\ead{d.trizna@pm.me}

\author[1]{Luca Demetrio}[]
\ead{luca.demetrio@unige.it}

\credit{Data curation, Writing - Original draft preparation}

\affiliation[2]{organization={Department of Computer, Control and Management Engineering, Sapienza University of Rome},
        addressline={Via Ariosto 25},
    city={Rome},
    postcode={00185}, 
    country={Italy}}

\author[3]{Battista Biggio}
\ead{battista.biggio@unica.it}

\affiliation[3]{organization={Department of Electrical and Electronic Engineering, University of Cagliari},
    addressline={Via Marengo 3}, 
    city={Cagliari},
    postcode={09123}, 
    country={Italy}}

\author[4]{Ivan Tesfai Ogbu}
\ead{ivan.tesfai@rina.org}

\affiliation[4]{organization={Rina Consulting S.p.A.},
    addressline={Via Antonio Cecchi 6}, 
    city={Genova},
    postcode={16129}, 
    country={Italy}}

\author[1, 3]{Fabio Roli}
\ead{fabio.roli@unige.it}

\cortext[cor1]{Corresponding author}

\begin{abstract}
As a result of decades of research, Windows malware detection is approached through a plethora of techniques.
However, there is an ongoing mismatch between academia -- which pursues an optimal performances in terms of detection rate and low false alarms -- and the requirements of real-world scenarios.
In particular, academia focuses on combining static and dynamic analysis within a single or ensemble of models, falling into several pitfalls like (i) firing dynamic analysis without considering the computational burden it requires; (ii) discarding impossible-to-analyze samples; and (iii) analyzing robustness against adversarial attacks without considering that malware detectors are complemented with more non-machine-learning components. 
Thus, in this paper we bridge these gaps, by investigating the properties of malware detectors built with multiple and different types of analysis. To do so, we develop \pipeline, a Windows malware detection pipeline sequentially leveraging both static and dynamic analysis, interrupting computations as soon as one module triggers an alarm, requiring dynamic analysis only when needed.
Contrary to the state of the art, we investigate how to deal with samples that impede analyzes, showing how much they impact performances, concluding that it is better to flag them as legitimate to not drastically increase false alarms.
Lastly, we perform a robustness evaluation of \pipeline. 
Counter-intuitively, the injection of new content is either blocked more by signatures than dynamic analysis, due to byte artifacts created by the attack, or it is able to avoid detection from signatures, as they rely on constraints on file size disrupted by attacks.
As far as we know, we are the first to investigate the properties of sequential malware detectors, shedding light on their behavior in real production environment.

\end{abstract}


\maketitle

\section{Introduction}

Due to the rapid evolution of threats and skills of malware developers, the detection of Windows malware is an on-going challenge that has kept revolutionizing itself for more than two decades.
To better understand the scale of this never-ending arms race, every week $\sim$ 7M\footnote{\url{https://virustotal.com/statistics}} Windows malicious programs are uploaded on cloud-based antivirus engines to be analyzed.
To exacerbate the issue, malware developers create several variants of the same malicious program to avoid detection by signature-based antivirus, which spot threats through unique indicators observed in the past and matched in analyzed samples. 
Thus, modern detection engines fully embrace the machine learning (ML) paradigm, by directly inferring the maliciousness of program from data, being also able to generalize across variants.

While we can only glimpse the architectures developed by industrial companies through minimally-detailed white papers~\citep{kaspersky, saxe2015deep, eset, avira, windefender}, academic research mostly focuses on creating models with the best trade-off between detections and false alarms~\citep{raff2018malware, trizna2024nebula, anderson2018ember, jindal2019neurlux, raff2020getting, gibert2020rise}, with a strong focus on the latter they are extremely costly to handle~\citep{hidden_cost_fpr_intrusion, fpreset_blog}.
To do so, state-of-the-art techniques focus on single or ensemble models that separately or jointly leverage static and dynamic analysis.
While the first one infer maliciousness from both the structure and the content of input samples, the second one requires programs to be executed or emulated inside an isolated environment.

However, dynamic analysis is costly, requiring a clean environment at each analysis where samples can be ``detonated'' to manifest their behavior. 
Every ML architecture that uses this kind of analysis must face this mandatory cost, and also plenty of proposed approaches consider features extracted with dynamic analysis fused or stacked together~\citep{dambra2023decoding, trizna2022quo} with the static ones.
This not only expands the complexity of the feature extraction phase, making it more costly and time-consuming, but also the overall predictive capability might not even benefit from this addition: \cite{dambra2023decoding} highlight a worrisome trend where dynamic analysis is neither better or complementary to the static one, diminishing the belief of tracing execution as the Swiss knife against malware.
On the other hand, most of the proposed techniques hide the crashes encountered by static analysis while extracting features, resolving the issue by discarding those impossible-to-analyze samples.
While this is not a problem at the research stage, such a behavior is not admissible in production environments where an answer must always be given to the users who requested an analysis.

To further exacerbate problems of static analysis that can be encountered in production, ML Windows malware detectors have been shown to be vulnerable to adversarial EXEmples~\citep{demetrio2021functionality, demetrio2021adversarial, anderson2017evading, lucas2021malware}, carefully-crafted programs tailored to evade detection.
These are constructed by manipulating the structure of samples, by either adding new or replacing existing content, thus interfering with the patterns learned at training time.
With almost no implementations available for attacks that target dynamic classifiers~\citep{rosenberg2018generic}, robustness against EXEmples is only computed against static detectors, thanks to reproducible open-source software~\citep{demetrio2021secmlmalware}.
However, while these attacks have only been tested against specific targets, they have not been evaluated against a production-ready Windows malware pipeline that comprises many different components, such as dynamic analysis or signature matching.
In theory, adversarial EXEmples against static detection should not have any effect on execution traces of tampered programs~\citep{demetrio2021adversarial, demetrio2021functionality}, being also more detectable by dynamic analysis, but no investigations have been conducted in this direction.

Hence, in this work we conduct a detailed investigation on the behavior of sequential malware detection pipeline, by addressing the highlighted roadblocks of the state of the art.
To do so, we leverage existing technology to build \pipeline, a malware detector that matches the needs of production environment.

\pipeline is built on three components: (i) pattern-matching with YARA rules to rapidly filter out known samples; (ii) static malware detection with state-of-the-art models to capture most of the threats; and (iii) dynamic analysis with emulation to fine-tune results on more difficult samples.
Input programs traverse the pipeline sequentially, halting the process at the first detection.
This step reduces the need for firing dynamic analysis, focusing its resource-demanding feature extraction process only when needed.
Static analysis is achieved with a mixture of models, combining end-to-end --less predictive, but resistant to crashes due to the absence of feature extraction-- and feature-extraction-based --more predictive, but likely to encounter pre-processing errors-- detection to reduce the number of crashes.
Through \pipeline, which mimics an industrial malware detector, we want to answer three research questions:

\mypar{RQ1:} how to properly deploy a malware detection software in presence of pre-processing errors at different stages?

\mypar{RQ2:} which are the performances of sequential analysis, compared to existing single and hybrid approaches?

\mypar{RQ3:} is it possible to tune the detection threshold of each component of \pipeline w.r.t. to a single validation set?

\mypar{RQ4:} how much detection performance and false positive rates varies when deploying calibrated pipeline w.r.t. the na\"ive combination of pre-trained models?

\mypar{RQ5:} what is the robustness of an end-to-end pipeline of pre-built components? 

\mypar{RQ6:} what is the overhead brought by a sequential pipeline in analysing input samples? 

To answer these questions, we provide an extensive experimental analysis conducted on two dataset, where we compare \pipeline to state-of-the-art models in terms of detection, false alarms and robustness.
Our findings highlight interesting trends, that can be summarised in the following take-home messages:

\mypar{Take-home message 1.} Counter-intuitively, samples causing errors during pre-processing should advance in the pipeline, being labeled as benign in case no modules are able to analyze them.
This reduces drastically the number of false alarms, and partially the detection, still being acceptable in production environments.

\mypar{Take-home message 2.} \pipeline outperforms all the competitors by keeping an extremely-low number of false positive rates, without discarding any sample in the process.
Also, \pipeline is much faster than all models leveraging dynamic analysis, launching it only when needed.

\mypar{Take-home message 3.} We show how to tune the detection threshold of each component of \pipeline by applying grid search, highlighting their substantial change w.r.t. to the original ones computed at their training time.

\mypar{Take-home message 4.} Counter-intuitively, calibrated thresholds of the components of \pipeline does not bring relevant improvements to accuracy and false positive rate.

\mypar{Take-home message 5.} \pipeline is more robust against static adversarial EXEmples, but not thanks to dynamic analysis.
We show that artifacts generated by attacks are detected by YARA rules that were not matching the unperturbed program.
We show that content injection attacks might have a marginal effect on functionality, since content is displaced and retrieved at different addresses at runtime, causing the dynamic-based ML model to reduce its score.
Lastly, we highlight that some signatures, relying on constraints on the filesize, are bypassed by adversarial attacks.

\mypar{Take-home message 6.} \pipeline does not negatively impact the time needed for analysing samples, since most of the malicious programs are rapidly recognized by static analysis, requiring the dynamic one only for benignware.

As far as we know, we are the first to investigate the behavior of sequential pipelines like \pipeline, shortening the gap between real production environment and academia. 
In particular, we are the first to perform a realistic testing, not only in regular industrial environment where pre-processing errors must be considered, but also in presence of skillful attackers that try to bypass all the layers of detection.

The rest of the paper is organized as follows: we firstly introduce the background concepts needed to understand our work (\autoref{sec:background}), followed by the implementation details of \pipeline (\autoref{sec:pipeline}).
We then continue by describing our experiments (\autoref{sec:exp_setup}), and which results we derive from them, along with the answers to our research questions (\autoref{sec:exp_results}).
We conclude the paper by detailing the limitations of our approach (\autoref{sec:limitations}), possible future research directions, and final remarks (\autoref{sec:conclusions}).

\section{Background}
\label{sec:background}
In this section, we introduce the main concepts and technologies that constitute the fundamentals for our research. 

\begin{figure}
    \centering
    \includegraphics[width=0.6\columnwidth]{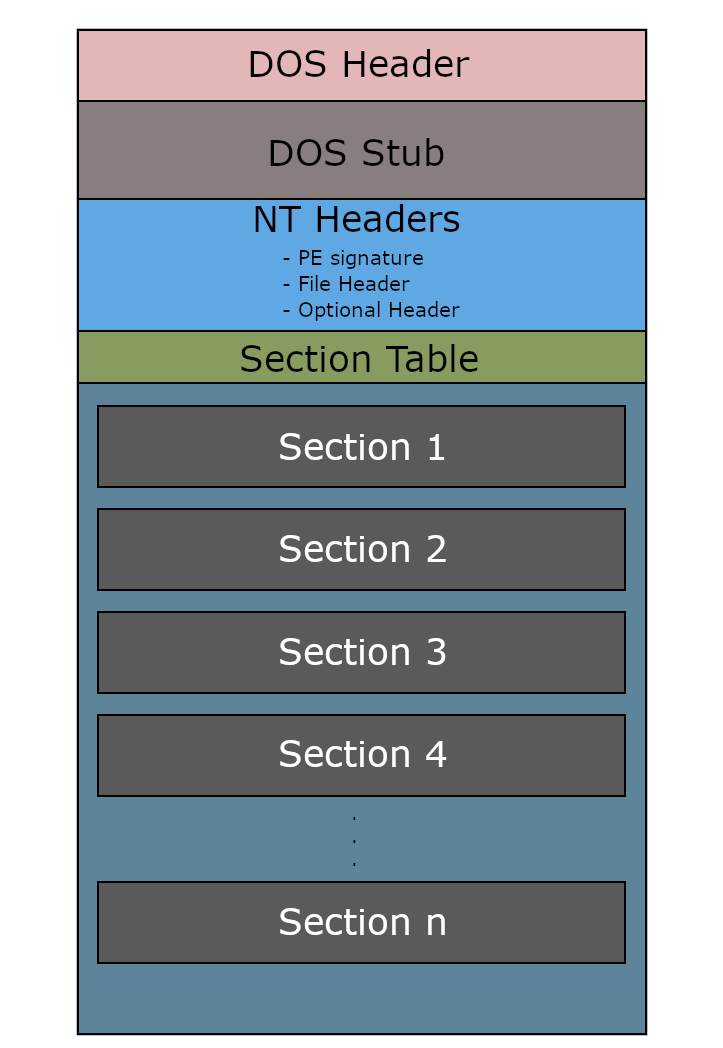}
    \caption{Graphical representation of the Windows PE format.}
    \label{fig:peformat}
\end{figure}

\mypar{Windows Portable Executable (PE) Format.} This is the standard format for Windows programs, used for both executables and shared libraries (DLLs) (\autoref{fig:peformat}).
It describes how files are stored on disk and how to be properly loaded in memory. 

\mysubpar{DOS Header + Stub.} These unused chunks of bytes represent a valid DOS program, kept for retrocompatibility.

\mysubpar{NT Headers.} These bytes represent the real header of the program, containing the \texttt{PE} signature and all the information needed by the OS to load the content in memory.

\mysubpar{Sections.} These contain code and assets of the program to load. Usually, the first section contains machine instructions of the software, followed by storages for initialized variables, resources, and other relevant information.

\mypar{Windows Malware Detection.}
To stop the spreading of malware, various techniques have been developed leveraging either static, dynamic, or both types of analysis.

\mysubpar{Static analysis.} This type of analysis is based on the extraction of relevant metrics from the structure and content of analyzed samples~\citep{raff2018malware, raff2020getting, anderson2018ember}, without the need for executing it.
The most na\"ive static analysis methodology is posed by pattern-matching with well-known signatures~\citep{YaraRules, capa}, but such a technique is not robust against all the malware variants that are released on a daily basis.
Thus, ML models are trained on static features to gain general performance also on unseen samples.

\mysubpar{Dynamic analysis.} This type of analysis concentrates on characterizing the behavior of programs~\citep{trizna2024nebula, jindal2019neurlux} by recording their trace of execution (spawned processes, API calls, reached websites, accessed registry keys) while being detonated inside an isolated environment, such as a sandbox, or through emulation.
After having collected all the events, these are pre-processed to be fed to machine learning models as training data.

\mysubpar{Hybrid analysis.} This type of analysis merges both static and dynamic information, often obtaining better results due to the greater amount of collected malware characteristics~\citep{trizna2022quo}.
This can be achieved by either stacking together information, or by fusing the representation in deep neural networks~\citep{gibert2020rise}.

\mypar{Adversarial EXEmples.} The rise of ML in malware detection brought in parallel the rise of Adversarial Machine Learning~\citep{yuan2019advrsarial, wiyatno2019adversarial, biggio2013evasion, biggio2018wild}.
In particular, in the domain of malware detection, this translates in the creation of adversarial EXEmples~\citep{demetrio2021adversarial, demetrio2021functionality}, carefully-perturbed programs tailored to fool ML-based detectors.
While limited research focuses on attacking dynamic detectors~\citep{rosenberg2018generic}, most of the efforts have been focused on evading static classifiers~\citep{demetrio2021adversarial, demetrio2021functionality, anderson2017evading, lucas2021malware}.
These attacks work by either replacing existing content or by injecting new one that disrupt the pattern learned at training time.

\section{\pipeline: Sequential Pipeline For Windows Malware Detection}
\label{sec:pipeline}
\begin{figure*}[t!]
\centering
  \includegraphics[width=\textwidth]{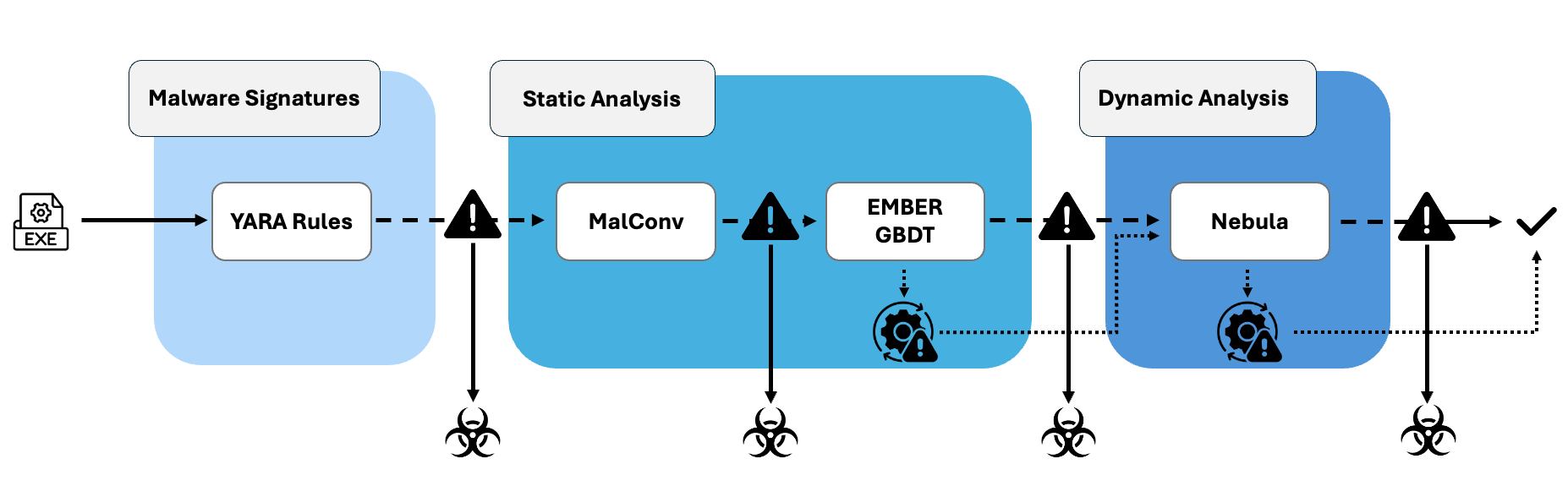}
  \caption{The architecture of \pipeline. Input traverses the pipeline by triggering each module sequentially, and computations are halted when one module raises an alert. In presence of pre-processing errors, \pipeline forwards the input to the next module. This can happen in non-end-to-end modules such as \gbdt and Nebula.}
  \label{fig:teaser}
\end{figure*}
We now describe \pipeline, a Windows malware detector built on different components, leveraging both static and dynamic analysis as depicted in \autoref{fig:teaser}.
Different from state-of-the-art techniques that use single or ensemble models, \pipeline performs predictions by sequentially testing different modules, halting computations for a specific sample when it is detected as malicious by one component.
Instead of merging together static and dynamic analysis that would require detonating samples in sandboxes, \pipeline leverages emulation to retrieve execution traces, and such operation is only required for programs that are flagged as non-malicious by all the previous modules.
In this way, we minimize the time of detection for malicious files, also thanks to the order of the modules, sorted from the fastest to the slowest.
When one of the modules fails to analyze a sample due to errors or crashes of pre-processing, \pipeline continues the analysis by passing the input to the next one.
If no module is able to successfully compute a prediction, we flag the sample as benign. 
We will later show that this counter-intuitive choice is applied to keep the number of false alarms extremely low, without reducing much the predictions on malicious samples (\autoref{sec:single_modules}).
We now detail all the components of \pipeline, by discussing their implementation and design choices.

\subsection{Pattern-matching Detection with Signatures}
The first module of \pipeline uses YARA\footnote{\url{https://github.com/VirusTotal/yara}}, which is a pattern-matching tool for detecting already-known malware from signatures.
These are textual descriptions containing binary patterns that identify a group of malicious programs, usually structured as shown in \autoref{lst:yara_rule}.
Each rule contains metadata used to describe its function, followed by strings or patterns (that can be hexadecimal strings, regular expressions, etc.) that will be looked for inside programs. 
Lastly, rules must define the firing condition, which describes how malicious activity is spotted. These are described as an if-then-else conditional block that describes the detection algorithm.

\begin{minipage}{0.9\linewidth}
\begin{lstlisting}[
    label={lst:yara_rule},
    basicstyle=\ttfamily,
    commentstyle=\color{gray},
    showspaces=false,
    showstringspaces=false,
    breakindent=1em,
    breaklines=true,
    % xleftmargin=1.8em,
    keepspaces=true,
    frame=single,
    caption={Example of YARA rule.},
    captionpos=b,
    aboveskip=10pt % Adjust the value as needed
    ]
rule rule_name : 
{
    meta:
        description = "Rule Description"
        author = "name"
        date = ""
    strings: 
        $a = {Hex Pattern}
        $b = {Regex}
    condition: 
        $a or $b
}
\end{lstlisting}
\end{minipage}

\subsection{Machine Learning Static Analysis}
The second module of \pipeline leverages two static analysis machine learning malware detectors: (i) \malconv~\citep{raff2018malware}, and end-to-end deep neural network without feature extraction; and (ii) \gbdt~\citep{anderson2018ember}, a gradient boosting decision tree trained on data pre-processed with hand-crafted features. 

\mypar{\malconv.} Proposed by~\cite{raff2018malware}, this model has been developed to learn maliciousness directly from bytes, by taking in input whole executable files and returning probability scores.
It is implemented as a convolutional neural network (CNN), starting from an embedding layer that encodes bytes inside a space where distances are defined. Input is then processed through a gated convolution layer, a global max-pooling layer, and a fully connected layer that computes the final prediction.
\malconv is trained on the state-of-the-art dataset EMBER~\citep{anderson2018ember}, and it takes in input the first 1 MB of each sample. In case input programs are shorter than this amount, they are padded with a special value.

\mypar{\gbdt.} Proposed by \cite{anderson2018ember}, this model leverages a Gradient Boosted Decision Tree (GBDT) trained on the state-of-the-art dataset EMBER~\citep{anderson2018ember}, containing features extracted from Windows Portable Executables (PEs). Features consist of eight groups, including features extracted after parsing the PE, and "format-agnostic" features, containing characteristics obtained without the parsing. 
The parsed features groups are: (i) \textit{general file information}, including the information obtained from the PE header; (ii) \textit{header information} taken from the headers, like the target machine, the target subsystem and so on; (iii) \textit{imported functions}, reporting the imported functions by library; (iv) \textit{exported functions} as a list; (v) \textit{section information}, comprehending the properties of each section. The format-agnostic features are a (i) \textit{byte histogram}, representing the normalized counts of each byte value within the file, a (ii) \textit{byte-entropy histogram} which accounts for the entropy of the byte distribution of the file, applying a sliding window over the binary and (iii) \textit{string information} taken from the printable strings inside the PE.

Each input sample is first analyzed by \malconv, and, if the model is not raising any alert, it is passed to \gbdt.
In this way, we first analyze samples with a faster model with no pre-processing, thus reducing the overall number of possible errors.

\subsection{Machine Learning Dynamic Analysis}
\label{sec:nebula}
The last module of \pipeline leverages dynamic analysis to trace the execution of samples, and it should help \pipeline to recognize obfuscated and packed samples that evaded the previous modules.
Among the recently released models for dynamic analysis, we select Nebula~\citep{trizna2024nebula}, a pipeline influenced by advances of Large Language Models (LLM)~\citep{radford2018improving} employing the self-attention neural mechanism to analyze dynamic analysis reports and classify samples.
Nebula employs Windows kernel emulation as a compromise between computational complexity and coverage. While emulation is cheaper than system virtualization, it is more prone to dynamic analysis errors.
Nebula introduces domain-knowledge-influenced filters to distill behavioral reports from redundant or irrelevant information like memory addresses or PE file segment hashes. Further, the behavioral report is processed by the Transformer encoder neural network, producing a probability of maliciousness.
We use pre-trained objects released by Nebula, trained on a public dataset~\citep{trizna2022quo}, comprising $\approx$75k samples collected in January 2022, spanning across seven malware types and $\approx$25k benignware samples.

\subsection{\pipeline Implementation}
We now present how we serialize the components we have described. 
\pipeline processes input programs by sequentially passing them from module to module, halting computations if one of those modules raises an alert. 
If one of the modules crashes due to pre-processing errors, \pipeline skips that module by passing the input to the next one. This can happen for EMBER-GBDT and Nebula, which require heavy pre-processing.
We discuss classification error management in \autoref{sec:exp_results}, by showing that it is better to label samples as benign in case of pre-processing crashes.

For the first YARA module, we collect 2.7k rules available in open GitHub repositories, accessed until November 2023\footnote{\url{https://github.com/bartblaze/Yara-rules/tree/master/rules}}\footnote{\url{https://github.com/elastic/protections-artifacts/tree/main/yara/rules}}\footnote{\url{https://github.com/malpedia/signator-rules/tree/main/rules}}\footnote{\url{https://github.com/Neo23x0/signature-base/tree/master/yara}}\footnote{\url{https://github.com/Yara-Rules/rules/tree/master/malware}}.

For the machine learning static analysis module, we leverage a library called \texttt{secml\_malware}~\citep{demetrio2021secmlmalware}\footnote{\url{ https://github.com/pralab/secml\_malware}} which wraps both \malconv and \gbdt models.
Instead of training both from scratch, we leverage pre-trained open-source implementations of both~\footnote{\url{https://github.com/endgameinc/malware_evasion_competition}}.
As the last module of \pipeline leverages the Nebula model provided by its original repository\footnote{\url{https://github.com/dtrizna/nebula}}. 
Emulation, which is the core of Nebula, is achieved through Speakeasy~\citep{speakeasy}, a Windows kernel emulation library, which generates a behavioral report in a JSON file.

\section{Experimental Setting}
\label{sec:exp_setup}
We now introduce all the experiments we will perform on \pipeline, and we will share results in \autoref{sec:exp_results}.

\mypar{Datasets.}
In our experiments, we employed two different datasets, that deferred in cardinality, malware families, benignware, and period of collections. 
The first one ($\dsone$) is composed of 5k malware, collected from VirusTotal before year 2018, and 3.5k benignware, harvested from GitHub and clean Windows installations.
We show its composition in \autoref{table:ds_1}, drafted again utilizing VirusTotal.
As regards malware families, this dataset is unbalanced, with predominance of two specific families among all the others.
While this composition might seem unsuitable for experiments, we believe it can be treated as a realistic snapshot of real production environments, where there is no control over the incoming threats to analyze.
\begin{table*}[ht]
    \centering
    \caption{Composition of $\dsone$ divided by malware class. There are 551 samples we were unable to correctly label.}
    \begin{tabular}{cccccccccccc}
        \toprule
        \textbf{Class} & \textbf{Benign} & \textbf{Backdoor} & \textbf{Downloader} & \textbf{Grayw.} & \textbf{Miner} & \textbf{Ransomw.} & \textbf{Roguew.} & \textbf{Spyw.} &\textbf{Virus} & \textbf{Worm} & \textbf{Unlabel.} \\
        \toprule
        \textbf{Count} & 3446 & 376 & 1354 & 1328 & 23 & 193 & 10 & 47 & 719 & 381 & 551\\
        \bottomrule
    \end{tabular}
    \label{table:ds_1}
\end{table*}
\begin{table*}[ht]
    \centering
    \caption{Composition of  $\dstwo$~\cite{trizna2022quo} by 
      malware class.}
    \begin{tabular}{ccccccccc}
        \toprule
        \textbf{Class} & \textbf{Benign} & \textbf{Backdoor} & \textbf{Coinminer} & \textbf{Dropper} & \textbf{Keylogger} & \textbf{Ransomw.} & \textbf{RAT} & \textbf{Trojan} \\
        \toprule
        \textbf{Count} & 10000 & 2500 & 2500 & 2500 & 2500 & 2500 & 2500 & 2500 \\
        \bottomrule
    \end{tabular}
    \label{table:speakeasy_dataset}
\end{table*}
The second dataset ($\dstwo$) reflects the test set of dynamic analysis records published by~\citep{trizna2022quo}, comprising 10K benign-ware and 17.5K malware samples spanned across seven types (such as ransomware, trojans, keyloggers, etc.) as shown in \autoref{table:speakeasy_dataset}. 
The dataset was collected in April 2022 by partnering with an undisclosed security vendor. 
We re-collected PE files from public data sources based on released hashes to perform static and YARA level analysis applicable only to raw PE bytes.
Contrary to $\dstwo$, this dataset has perfect balances between malware families, making it a good baseline for fairly assessing performances across different techniques.

\mypar{Evaluation metrics.} We compute several metrics to characterize the performance of \pipeline and baseline models:

\mysubpar{True Positive Rate (TPR)} to evaluate the module/model capability to correctly label effective malware samples, i.e., detect malicious PEs.  

\mysubpar{False Positive Rate (FPR)} to evaluate the rate of false alarms raised by each model/module. This is a key parameter in a production scenario, where high FPR compromises the usability of a malware detector.

\mysubpar{F1-Score (F1)} to evaluate whether \pipeline achieves high performances without excessive false positives or negatives.

\mysubpar{Mean Detection Time (MDT)} to evaluate the mean time that each component needs to analyze an input sample. We compute this metric on a subset of our datasets. 

\mysubpar{Error Rate (ER)} of each component to estimate the percentage of impossible-to-analyze samples. 
We provide an experimental explanation of the best way to label those samples, as a decision must be taken when deployed.

\mysubpar{Adversarial Detection Rate (ADR)} to assess robustness evaluation simply stating how many adversarial EXEmples are correctly recognized as malware after the attack.

\mypar{Comparison between \pipeline and single models.}
In the first experiment, we test the predictive capabilities of \pipeline on our two datasets, comparing it against each separate module.
In doing this we motivate the choice of our proposed architecture, especially for the dynamic analysis module. We show that the chosen Nebula is better than a competitor model, named Neurlux~\citep{jindal2019neurlux}, a LSTM model trained on dynamic reports extracted from emulation.
Also, we improve the comparison by including Quo.Vadis~\citep{trizna2022quo} which is a hybrid-analysis model that merges both static and dynamic features.
On the contrary to \pipeline, Quo.Vadis introduces a monolithic structure with single forward- and back-propagation path, processing static and dynamic components simultaneously, further employing a "meta-model" that accumulates representations from both analysis types to classify Windows executable samples.

\mypar{Effect of dynamic analysis in \pipeline Classification.}
Similarly to \citep{dambra2023decoding}, we investigate the efficacy of dynamic analysis. 
Different from their setting, here we use emulation as last step of the analysis without concatenating together both static and dynamic features.
Thereafter, we test the performances of \pipeline in two cases: in the former, we do not include the dynamic module, and in the latter, we test the full pipeline, including the model chosen after the analysis of the two separate models. 
Lastly, we also include \pipeline performance on detection divided per malware family contained in $\dsone$ and $\dstwo$.

\mypar{Calibration of the components of \pipeline.}
We now introduce the technique we adopt to calibrate decision thresholds of the modules of \pipeline. 
We divide our two datasets \dsone and \dstwo in two subsets, in which families are equally distributed: a validation set and a test set. 
In order to perform a fair calibration, we take half of the benignware samples from both \dsone and \dstwo as well. 
We perform a simple version of an exhaustive grid search with this methodology: our hyperparameters in the grid are the three decision thresholds of the three ML modules in \pipeline (namely, \malconv, \gbdt and Nebula). For each combination, we calculate the F1-score on the validation set and take the threshold combination with the best F1-score. We repeat this process with different cardinalities of the validation set, aiming at picking the best strategy in terms of the F1-score and remaining test set dimension. 
We take the same research space for each decision threshold: $\mathcal{R} = \{0.44, 0.46, \dots ,0.96, 0.98\}$.
Different from cross-validation, we do not re-train all the modules for each combination considering different folds, but we only evaluate scores on a validation set to properly tune decision thresholds. 
This methodology allows us to fairly compare the single modules and other state-of-the-art models with \pipeline, since we calibrate them too to the same FPR score as achieved by the "tuned" pipeline.

\mypar{Effect of Calibration on \pipeline Classification.}
We perform the exhaustive grid search as described in the previous section, first halving our datasets into a validation and a test set. We compute the F1-score for each thresholds combination on the validation set and we repeat the process taking other percentages of the total amount of samples at our disposal. We analyze if calibration results significantly change increasing the number of samples used for calibration. After choosing the best split strategy and the best decision thresholds, we test the "calibrated" \pipeline on the remaining test set. We discuss obtained results comparing the "non-calibrated" pipeline performance to assess how much advantage we can obtain from a (simple) threshold-tuning procedure. For the sake of completeness, using the same validation set, we also select a threshold for every single model under test, picking the one that gives the same FPR as the calibrated \pipeline. With those thresholds, we repeat the same experiments with the single models. 

\mypar{Detection Time Analysis.}
Here, we want to compare the mean speed of each separate module or model, aiming to analyze the advantages and disadvantages in terms of time. For the sake of completeness, we report the technical characteristics of the machine we used for this analysis: we use a workstation equipped with an Intel® Xeon(R) Gold 5420, two Nvidia L40 GPUs, and 540 GB of RAM.
We calculate the MDT metric on two subsets of $\dsone$ and $\dstwo$. The first one comprises 500 malware samples from $\dsone$ and 500 malware samples from $\dstwo$. Similarly, we take 500 benignware samples from the two datasets. We also compute the standard deviation to show how the detection time can vary. For this evaluation, we separate malware and benignware samples because they differ in file dimension: benignware are usually bigger than malware, and this enlarges emulation, processing and classification times. Moreover, benign samples always pass through all \pipeline's modules by design, and we prefer to evaluate times separately. To differ the two metrics, we denote with MDT$_m$ and MDT$_g$ the calculation of MDT for malware and benignware subsets respectively.
Also, we compute detection times of \pipeline after calibration, since setting different thresholds impacts on where (by which module) a sample is detected.

\mypar{Pipeline Robustness Evaluation.}
We test \pipeline in terms of robustness, evaluating this metric with and without thresholds calibration.
Due to the complexity of landing end-to-end attacks against pipelines like \pipeline, we opt for computing \textit{transfer attacks}, by optimizing adversarial EXEmples~\citep{demetrio2021adversarial} -- carefully-manipulated Windows programs aimed to evade detection -- against a surrogate detector with similar capabilities, and then test them against \pipeline.
Thus, we envision an attacker targeting the \gbdt model (which proves to be the best performing in most scenarios in terms of TPR and FPR, as we are going to discuss in \autoref{sec:exp_results}), knowing is used inside the pipeline.
To do so, we rely on GAMMA, a state-of-the-art black-box attack~\citep{demetrio2021functionality} that injects non-executable portion of legitimate programs into malware.
GAMMA leverages a genetic algorithm to select content that will included through either (i) the creation of new sections or (ii) appending bytes at the end (padding).
GAMMA iteratively requests classification of perturbed samples to optimize the amount of injected content, until either evasion is achieved or the query budget is terminated.

Thus, we consider 1k samples classified as malicious by \gbdt from $\dstwo$, and, as content to inject, we consider 75 benign sections picked from $\dstwo$ benignware samples. 
Since GAMMA relies on regularization to control the total filesize of adversarial EXEmples, we set its parameter $\lambda = 1\times 10^{-6}$, and we set the maximum number of queries to 100.
These sections contain read-only data used by programs at runtime, but no further code is included inside adversarial EXEmples.

\section{Experimental Results}
\label{sec:exp_results}
We now describe all the findings we gathered from the experiments described in \autoref{sec:exp_setup}. 
To highlight the effect of different error-handling strategies, we will report in tables performance when flagging impossible-to-analyze samples both as malicious in brackets, and as benignware otherwise.

\subsection{Single Modules Classification}
\label{sec:single_modules}
\begin{figure}
    \centering
    \includegraphics[width=1\columnwidth]{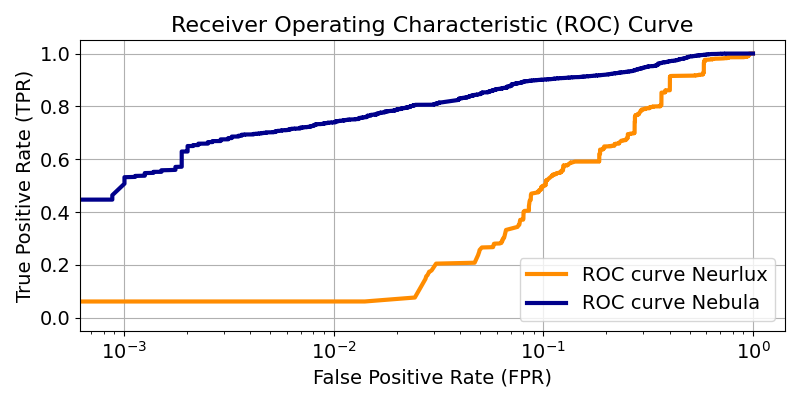}
    \caption{ROC curve of Nebula and Neurlux, tested on $\dstwo$. To compute these curves we discard impossible-to-analyze samples.}
    \label{fig:roc nebula neurlux}
\end{figure}

We test the performance of all the single modules with both $\dsone$ and $\dstwo$ datasets in our possession, and we present results of each module in \autoref{separate module test}.
We report the values of the described metrics when modules consider input as benignware in presence of pre-processing errors, by also including in brackets metrics computed considering classification errors as malware, as mentioned earlier in this section. 
\begin{table}[]
\caption{Performance of single \textbf{not-calibrated} models and \pipeline. Results in brackets are calculated considering detection errors as malware, and benignware otherwise.
We report the True Positive Rate (TPR), the False Positive Rate (FPR), F1 Score (F1), and the Error Rate (ER).}
\resizebox{\linewidth}{!}{%
\begin{tabular}{@{}ccccc@{}}
\toprule
\multicolumn{5}{c}{$\dsone$} \\ \toprule
\textbf{Model} & \textbf{TPR} & \textbf{FPR}  & \textbf{F1} & {\textbf{ER}} \\ \toprule
Signatures   &  0.15  & 5.2 $\times 10^{-3}$ & 0.27 & 0\%  \\
\malconv      &  0.78  & 2.0 $\times 10^{-2}$ & 0.87 & 0\% \\
GBDT         &  0.91 (0.92)  & 4.1 $\times 10^{-3}$ (0.12) & 0.95 (0.91) & 5.4\%  \\
Nebula     & 0.23 (0.61) & 2.3 $\times 10^{-2}$ (0.37) & 0.37 (0.64) & 39\%  \\
Quo.Vadis    & 0.64 (0.73) & 1.5 $\times 10^{-2}$ (0.25) & 0.78 (0.77) & 15\% \\
\midrule
\pipeline No Dyn.  & 0.94 (0.95) & 2.5 $\times 10^{-2}$ (0.14) & 0.96 (0.92) & 5.3\% \\
\pipeline & 0.94 (0.97) & 4.1 $\times 10^{-2}$ (0.38) & 0.95 (0.87) & 19\% \\ 
\toprule
\multicolumn{5}{c}{$\dstwo$} \\ \toprule
\textbf{Model} & \textbf{TPR} & \textbf{FPR} & \textbf{F1}  & \textbf{ER}                      \\ \toprule
Signatures   &  0.34  & $8.9\times10^{-3}$  & 0.51  & 0\% \\
\malconv      &  0.74  & $4.8\times10^{-2}$  & 0.84  & 0\% \\
GBDT         &  0.75 (0.75)  & 2.8 (3.6) $\times10^{-3}$ & 0.85 (0.85) &  0.036\% \\
Nebula  & 0.41 (0.86) & 7.8 $\times 10^{-3}$ (0.21) & 0.58 (0.87)  & 45\%  \\ 
Quo.Vadis    & 0.80 (0.84) & 9.2 $\times 10^{-3}$ (8.0 $\times 10^{-2}$) & 0.89 (0.89) & 4.8\% \\ 
\midrule
\pipeline No Dyn.  & 0.88 (0.88) & 5.8 (5.9) $\times 10^{-2}$ & 0.91 (0.91)  & 0.029\% \\
\pipeline & 0.89 (0.96) & 6.4 $\times 10^{-2}$ (0.26) & 0.93 (0.91)  & 20\%\\ 
\bottomrule
\end{tabular}}
\label{separate module test}
\end{table}
To better clarify the choice of the dynamic malware detector, we compare two state-of-the-art models: (i) Nebula, the transformer architecture described in \autoref{sec:nebula}; and Neurlux~\citep{jindal2019neurlux}.
We plot the ROC curves of the two models, considering only successfully classified samples belonging to \dstwo, dropping the ones that resulted in classification errors. 
We see that Nebula outperforms Neurlux at each FPR, as shown in \autoref{fig:roc nebula neurlux}, that is the reason for our choice in the development of the dynamic module.
Also, we want to calibrate our models finding a threshold at 1\% FPR if possible, but Neurlux achieves very-low TPR at that threshold, which is unacceptable in a production detector.
Thus, coherently with previous work~\citep{trizna2024nebula}, we discarded Neurlux as possible module for \pipeline.

As anticipated, we also include in the comparison the hybrid model Quo.Vadis~\citep{trizna2022quo}, considering that its approach of static and dynamic feature fusion is widespread in malware detection research. 
We compute metrics for Quo.Vadis considering a 1\% FPR threshold computed on $\dstwo$. 
We specify that stock \malconv and \gbdt models have a decision threshold computed at 1\% FPR on their original test sets, while for Nebula we take the 1\% FPR threshold from the ROC in \autoref{fig:roc nebula neurlux}.

Before drawing a conclusions on the predictive capabilities of tested models, we note that the best in terms of TPR, FPR, and F1 are achieved by considering pre-processing errors as benignware.
Thus, we are able to properly answer \textbf{RQ1} for single models, by counter intuitively asserting that, due to the majority of impossible-to-analyze samples among benignware, it is better to slightly decrease TPR to favour a low FPR regime.

By looking at the best trade-off between FPR and TPR, we can not isolate a clear winner among models evaluated singularly. 
In particular, both \gbdt and Quo.Vadis achieve the best performance on both datasets, with extremely low false positive rates and good TPR.
It is worth to notice that Quo.Vadis also incorporates the same feature set provided by~\citep{anderson2018ember}, which is also used for \gbdt.
However, if we also include the error rate (ER) as part of our analysis, we can observe that \gbdt is clearly the winner of this analysis, with a very low percentage of crashes on both datasets.
On the contrary, Quo.Vadis achieves a peak of 15\% of pre-processing errors due to the combination of both static feature extraction and emulation.
As competing candidate to Quo.Vadis, also \malconv shows matching performance with these models, by also being characterized with zero pre-processing errors thanks to it end-to-end structure.

The best technique in terms of FPR is pattern-patching samples with YARA rules, but, as expected, the TPRs are very low for both datasets.
In numbers, our sample dataset of rules comprising 2.7k signatures is only able to detect 750 malware in $\dsone$ and 6k in $\dstwo$, scoring the least predictive detector.
While this analysis highlights that, on average, each rule detects roughly 22 samples of $\dstwo$, it is clear that YARA rules might match other models' predictive capabilities only with a larger set of signatures.
Lastly, dynamic analysis with emulation has a non-negligible number of errors, which hinders its already low metrics and its real impact in a multi-stage detector, while also inflating the number of false positives or negative depending on the error-handling strategy we consider.
This finding might be correlated with the fact that modeling the behavior of malware in an end-to-end perspective is a daunting task, thus confirming the superiority of static analysis~\citep{dambra2023decoding}. 

\begin{tcolorbox}[colback=red!40, colframe = black, boxrule=0.5pt]
\textbf{Take-home message 1:} The best error management policy for detection errors is to consider errors as benignware, especially for a low FPR, crucial in production to avoid false alarms. Purely-static model GBDT gives the best compromise between TPR, FPR, and ER, with respect to the others. Nebula struggles in being competitive due to the high ER, while Quo.Vadis demonstrates that is possible to improve the weakness of dynamic analysis by adding EMBER static features. 
\end{tcolorbox}

\subsection{\pipeline Classification Performance}
We now analyze the performances of \pipeline, with and without (\pipeline No Dyn.) the dynamic module as last step for the detection pipeline. 
This comparison will grant us insights on how much dynamic analysis improve the quality of predictions when used as a sequential detection mechanism, and we show the results in \autoref{separate module test}.
For both settings compared to single modules, we clearly notice a drastic improvement in the trade-off between TPR and FPR for both datasets.
While it is true that \gbdt achieves lower FPR, it is worth noticing that \pipeline increases TPR by 3\% and 13\% on $\dsone$ and $\dstwo$ respectively.
This increment is given by the sequential order of techniques of our methodology, that patches blind spots of single strategies thanks to the multitude of increasingly-accurate models.
Thus, ER is lower compared to the dynamic module alone in both scenarios for two main reasons: (i) fewer malware samples that caused emulation to crash proceed to the last step of the pipeline, since they are stopped by earlier modules, and (ii) most of the pre-processing crashes are caused by benignware.
Thus, we complete the answer to \textbf{RQ1}, by stating that also for \pipeline it is better to treat impossible-to-analyze samples as benign, to avoid higher FPR (which is 26\% in the case of \pipeline Dyn.).
\begin{table*}[ht]
    \caption{True Positive Rate (TPR) of \textbf{non-calibrated} \pipeline divided by malware family inside in $\dsone$. Results in brackets are calculated considering detection errors as malware. They are considered benignware otherwise.}
    \centering
    \begin{tabular}{cccccccccc}
        \toprule
        \textbf{Architecture} & \textbf{Backdoor} & \textbf{Downloader} & \textbf{Grayw.} & \textbf{Miner} & \textbf{Ransomw.} & \textbf{Roguew.} & \textbf{Spyware} &\textbf{Virus} & \textbf{Worm}  \\
        \toprule
        \textbf{\pipeline No Dyn.} & 0.97 (0.97) & 0.97 (0.98) & 0.93 (0.93) & 1.0 (1.0) & 0.96 (0.96) & 0.90 (0.90) & 0.94 (0.94) & 0.99 (0.99) & 0.99 (0.99)  \\
        \textbf{\pipeline} & 0.97 (0.98) & 0.97 (0.99) & 0.93 (0.96) & 1.0 (1.0) & 0.97 (0.98) & 0.90 (1.0) & 0.94 (1.0) & 0.99 (0.99) & 0.99 (0.99)  \\
        \bottomrule
    \end{tabular}
    \label{table:family_performance_ds1}
\end{table*}
\begin{table*}[ht]
    \caption{True Positive Rate (TPR) of \textbf{non-calibrated} \pipeline divided by malware family inside in $\dstwo$. Results in brackets are calculated considering detection errors as malware. They are considered benignware otherwise.}
    \begin{tabular}{cccccccc}
        \toprule
        \textbf{Architecture}  & \textbf{Backdoor} & \textbf{Coinminer} & \textbf{Dropper} & \textbf{Keylogger} & \textbf{Ransomw.} & \textbf{RAT} & \textbf{Trojan} \\
        \toprule
        \textbf{\pipeline No Dyn.} & 0.99 (0.99) & 0.97 (0.97)& 0.83 (0.83) & 0.43 (0.43) & 0.95 (0.95) & 0.98 (0.98) & 0.99 (0.99) \\
        \textbf{\pipeline} & 0.99 (0.99) & 0.97 (0.99) & 0.83 (0.99) & 0.52 (0.81) & 0.97 (0.97) & 0.98 (0.99) & 0.99 (0.99)  \\
        \bottomrule
    \end{tabular}
    \label{table:family_performance_ds2}
\end{table*}
As for differences between the pipeline with the dynamic module and without, we can see a negligible improvement in TPR and only in with $\dstwo$ at the expense of FPR for both datasets. 

Also, to better understand such an improvement, we display the distinct predictive capabilities of both strategies on the different families that compose our datasets, by showing results in \autoref{table:family_performance_ds1} for $\dsone$ and \autoref{table:family_performance_ds2} for $\dstwo$.
For the first dataset, which is severely imbalanced in terms of malware families, we can notice an identical detection rate when treating errors as benignware.
Also, the number of errors is very low, with a peak of 3\% (40 samples) on Grayware programs, there is no clear advantage brought by dynamic analysis on this dataset.

Similarly, for the second dataset, both techniques score identical results, with low TPR for Dropper and Keylogger families.
Furthermore, this trend is confirmed by the F1-Score performed by the pipeline.
On this dataset, we confirm what has already been stated by~\citep{dambra2023decoding}, that dynamic analysis marginally helps in detecting specific families that are harder to detect by static analysis.
However, by also looking at numbers, all the emulation errors are also focused on those families as well.
Thus, it is possible that these numbers would further improve with a better emulation or virtualization tool as back-end for the dynamic analysis module.
We derive that adding a dynamic module at the end of a sequential pipeline for malware detection might not improve performances as much as expected.
In particular, this module might increase the errors encountered at analysis time, which must be dealt with accordingly.
Secondly, different from Dambra et al. which consider models with static and dynamic features concatenated together, dynamic analysis as a separate part of a detection pipeline does not increase the capabilities of the overall system.

Hence, we can answer \textbf{RQ2}, by stating the sequential architecture of \pipeline improves TPR within a low FPR regime than single models alone, but not thanks to the dynamic module. 
\begin{tcolorbox}[colback=red!40, colframe = black, boxrule=0.5pt]
    \textbf{Take-home message 2:} \pipeline outperforms single models in terms of the considered metrics, by also reducing the number of pre-processing errors.
    We complement previous findings~\citep{dambra2023decoding} by highlighting the minimal increment in performance brought by dynamic analysis in binary classification. 
\end{tcolorbox}

\subsection{\pipeline Performance After Calibration}
Here we present the calibration process of \pipeline, and how it changes the performance of all the considered models.

\begin{figure}
    \centering
    \includegraphics[width=1\columnwidth]{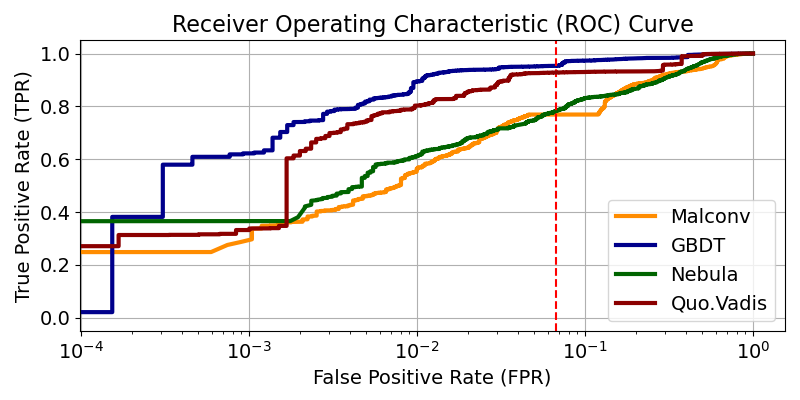}
    \caption{ROC curve of all the models under test. We take a decision threshold for each model fixing FPR at 6.7\% as scored in the validation process, to have a fair comparison with \pipeline.}
    \label{fig:roc_on_validation}
\end{figure}

\mypar{Performance on Validation Set.}
As described in \autoref{sec:exp_setup}, we perform an exhaustive grid search to select the best combination of decision thresholds characterizing \pipeline's models.
We pick thresholds from three different sets of values, one for each ML module inside our pipeline. Then, we compute TPR, FPR and F1-score of \pipeline with a validation set.
We adopt different splitting strategies to extract the latter from our data, and we chose the thresholds that maximise the F1-score of \pipeline, and we show results in \autoref{validation_scores}.
As the table suggests, there is no general improvement by varying the size of the validation set.
Thus, we take the 50\% splitting strategy since we can test the found thresholds combinations with a more representative and bigger test set even though the absolute winner in terms of F1-score is the 80\% strategy.
The optimal thresholds vary substantially from those determined at a 1\% FPR on the original datasets of the models. \malconv threshold is increased from 0.5 to 0.76, \gbdt threshold is significantly decreased from 0.82 to 0.44 and Nebula threshold is lowered from 0.97 to 0.70.
Hence, we can conclude that \malconv is mostly used by \pipeline to detect malware with high-confidence, leaving samples with lower score to \gbdt (whose threshold has been halved).
Thanks to this calibration, we can select the same FPR for all the models in our analysis, to produce a fair comparison.
Thus, we compute ROC curves for the other models and stand-alone modules to find thresholds achieving the same FPR, as depicted in \autoref{fig:roc_on_validation}).
In the remaining part of this subsection, we use found thresholds for all models and \pipeline to compare results in a calibrated and fair experimental setting.
\begin{tcolorbox}[colback=red!40, colframe = black, boxrule=0.5pt]
    \textbf{Take-home message 3}: All the detection thresholds of \pipeline can be fine-tuned on a validation set with a simple grid-search, substantially changing the behavior of each module differently than when used standalone.
\end{tcolorbox}
\begin{table}[]
\caption{Performance of \pipeline and optimal threshold computed on the validation set based on the best F1-Score. We report different splits of the union of \dsone and \dstwo, and the correspondent best thresholds for \malconv, \gbdt, and Nebula.}
\begin{tabular}{@{}ccccccc@{}}
\toprule
\textbf{Validation Split} & \textbf{TPR} & \textbf{FPR} & \textbf{F1-Score} & \textbf{$\mathcal{T}_M$} & \textbf{$\mathcal{T}_E$}& \textbf{$\mathcal{T}_N$}\\ \midrule
\textbf{50\%}             & \textbf{0.962}        & \textbf{0.067}        & \textbf{0.955} & \textbf{0.76} & \textbf{0.44} & \textbf{0.70}  \\
60\%                      & 0.961        & 0.064        & 0.955   & 0.76 & 0.44 & 0.70         \\
70\%                      & 0.960        & 0.067        & 0.953   & 0.76 & 0.50 & 0.70          \\
80\%                      & 0.961        & 0.063        & 0.956   & 0.76 & 0.5 & 0.72          \\ \bottomrule
\end{tabular}
\label{validation_scores}
\end{table}

\mypar{Performance on Test Set.}
We now analyze results on the test set, dividing metrics by dataset as we did in \autoref{sec:single_modules}.
Results are reported in \autoref{testset_results}. We compare the calibrated \pipeline with other models, setting their thresholds to the same FPR value as reported by our pipeline. 
We can see that on \dsone we have the same results as in the "non-calibrated" scenario, while on \dstwo, our pipeline reports higher TPR and lower FPR. 
This results is mainly due to \dstwo being more populated.
Hence, the validation set contain more representative samples of each of the contained malware families.
However, we can empirically state that calibrating with unbalanced datasets does not worsen results on the less present one while improving performance on the most prominent.
After that, we report another effect of the calibration: on \dstwo, removing dynamic analysis has a bigger impact than before, since Nebula is part of the process of fine-tuning. 
As regards comparisons with single models, \gbdt confirms to be the best performing one, even at this FPR, and our pipeline performs slightly worse than it in terms of TPR but with lower FPR. 
The error rates on the test set indicate a significant number of samples that are impossible to analyze when emulation is part of the analysis (as with Nebula, Quo.Vadis, and integral SLIFER). 
On the contrary, we encounter fewer errors for \gbdt and \pipeline without dynamic analysis. 
As ultimate test, F1-Score follows the trends already discovered earlier in this Section as well as in \autoref{separate module test} using stock thresholds.
Therefore, these results are consistent with our earlier findings.
Calibration highlights as well that the best error management policy is to consider analysis errors as benign to ensure a usable FPR.

\mypar{Results on testset families.}
For the sake of completeness, we report TPR on the test set for each malware family, and we report results in \autoref{table:family_performance_ds1_ts} and \autoref{table:family_performance_ds2_ts}.
Results are substantially aligned with the non-calibrated \pipeline. 
The calibration process improves detection of some of the malware families (especially the ones belonging to \dstwo), like Dropper and Ransomware, while under-performing on the Rogueware family, belonging to \dsone.
\begin{tcolorbox}[colback=red!40, colframe = black, boxrule=0.5pt]
    \textbf{Take-home message 4}: The calibration of detection thresholds of \pipeline can lead to more accurate results and better robustness in general, but its impact is negligible. 
\end{tcolorbox}

\begin{table}[]
\caption{Performance of \textbf{calibrated} single models and \pipeline.
Results in brackets are calculated considering detection errors as malware. They are considered benignware otherwise. We report the True Positive Rate (TPR), the False Positive Rate (FPR), F1 Score (F1), and the Error Rate (ER).}
\resizebox{\linewidth}{!}{
\begin{tabular}{@{}ccccc@{}}
\toprule
\multicolumn{5}{c}{$\dsone$} \\ \toprule
\textbf{Model} & \textbf{TPR} & \textbf{FPR}  & \textbf{F1} & {\textbf{ER}} \\ \toprule
Signatures   &    0.15&   7.0$\times10^{-3}$ & 0.26 & 0\% \\
\malconv      &  0.79 & 2.7$\times10^{-2}$& 0.87 &0\%\\
GBDT         &  0.96 (0.97)& 5.0$\times10^{-2}$ (0.17) & 0.96 (0.93) &5.6\%\\
Nebula     & 0.36 (0.70)& 6.0$\times 10^{-2}$ (0.37) & 0.51 (0.71) & 35\%\\
Quo.Vadis    & 0.87 (0.97)& 4.9$\times 10^{-2}$ (0.29) & 0.92 (0.89) & 15\%\\
\midrule
\pipeline No Dyn.  & 0.94 (0.95)& 2.3$\times 10^{-2}$ (0.15) & 0.96 (0.93) & 5.5\%\\
\pipeline & 0.94 (0.98)& 5.2$\times 10^{-2}$ (0.40) & 0.95 (0.87) & 16\%\\ 
\toprule
\multicolumn{5}{c}{$\dstwo$} \\ \toprule
\textbf{Model} & \textbf{TPR} & \textbf{FPR}  & \textbf{F1} & \textbf{ER} \\ \toprule
Signatures   &  0.35 & 9.6$\times10^{-3}$& 0.51 &0\%\\
\malconv      &  0.76& 5.7$\times10^{-2}$& 0.85 &0\%\\
GBDT         &  0.95 (0.95)& 7.1 (7.2)$\times10^{-2}$& 0.96 (0.96) &0.058\%\\
Nebula  & 0.47 (0.92)& 4.2$\times10^{-2}$ (0.25)& 0.63 (0.89) & 36\%\\ 
Quo.Vadis    & 0.89 (0.93)& 6.5$\times10^{-2}$ (0.14) & 0.93 (0.93)  & 4.7\%\\ 
\midrule
\pipeline No Dyn.  & 0.89 (0.89)& 4.7 (4.8)$\times10^{-2}$ & 0.93 (0.93) & 0.044\%\\
\pipeline & 0.92 (0.98)& 6.1$\times10^{-2}$ (0.27) & 0.94 (0.92) & 11\% \\ 
\bottomrule
\end{tabular}}
\label{testset_results}
\end{table}
\begin{table*}[ht]
    \caption{True Positive Rate (TPR) of \textbf{calibrated} \pipeline divided by malware family inside in $\dsone$. Results in brackets are calculated considering detection errors as malware. They are considered benignware otherwise.}
    \centering
    \begin{tabular}{cccccccccc}
        \toprule
        \textbf{Architecture} & \textbf{Backdoor} & \textbf{Downloader} & \textbf{Grayw.} & \textbf{Miner} & \textbf{Ransomw.} & \textbf{Roguew.} & \textbf{Spyware} &\textbf{Virus} & \textbf{Worm}  \\
        \toprule
        \textbf{\pipeline No Dyn.} & 0.98 (0.98)  & 0.96 (0.98) & 0.93 (0.93) & 1.0 (1.0) & 0.99 (0.99) & 0.80 (0.80) & 0.96 (0.96)& 0.99 (0.99)& 0.99 (0.99)\\
        \textbf{\pipeline} & 0.98 (0.99)& 0.96 (0.99)& 0.94 (0.97)& 1.0 (1.0)& 0.99 (1.0)& 0.80 (1.0)& 0.96 (1.0)& 0.99 (1.0)& 0.99 (0.99)\\
        \bottomrule
    \end{tabular}
    \label{table:family_performance_ds1_ts}
\end{table*}
\begin{table*}[ht]
    \caption{True Positive Rate (TPR) of \textbf{calibrated} \pipeline divided by malware family inside in $\dstwo$. Results in brackets are calculated considering detection errors as malware. They are considered benignware otherwise.}
    \begin{tabular}{cccccccc}
        \toprule
        \textbf{Architecture}  & \textbf{Backdoor} & \textbf{Coinminer} & \textbf{Dropper} & \textbf{Keylogger} & \textbf{Ransomw.} & \textbf{RAT} & \textbf{Trojan} \\
        \toprule
        \textbf{\pipeline No Dyn.} & 0.99 (0.99)& 0.98 (0.98)& 0.92 (0.92)& 0.42 (0.42)& 0.97 (0.97)& 0.99 (0.99)& 0.98 (0.99)\\
        \textbf{\pipeline} & 0.99 (0.99)& 0.98 (0.99)& 0.92 (1.0)& 0.57 (0.87)& 0.99 (0.99)& 0.99 (1.0)& 0.99 (0.99)\\
        \bottomrule
    \end{tabular}
    \label{table:family_performance_ds2_ts}
\end{table*}

\subsection{Robustness Against Transfer Attacks}

We now answer \textbf{RQ5} by testing the robustness of our architecture against adversarial EXEmples computed on \gbdt (with its "default" decision threshold at 1\% FPR) and transferred again the \pipeline, and we show results of GAMMA section-injection and padding in \autoref{adversarial_results_sections} and \autoref{adversarial_results_padding} respectively.
For the sake of clarity, we split this analysis into two paragraphs.

\mypar{GAMMA Section Injection Attack.}
From the randomly-sampled 1k malicious samples, GAMMA is able to produce 388 EXEmples that bypass \gbdt detection.
We use this set to compute the robustness against transfer attacks of \pipeline with and without the dynamic module.
Since we are interested also in the effect that signatures play in countering malware, we want to quantify how much they can be useful when dealing with adversarial EXEmples.
Thus, we also add to the comparison a version of \pipeline where YARA pattern-matching is removed (\pipeline No Sign.).
Lastly, to understand whether an attack that manipulate only the structure of a program might also affect a behavioral model, we also test the robustness of Nebula alone.
\begin{table}[]
\caption{Adversarial Detection Rate (ADR) of diverse architectures/model under \textbf{GAMMA Section Injection} attacks. We compare not manipulated malware samples and the same samples after manipulation (Adversarial EXEmples), considering \textbf{stock} and \textbf{calibrated} architecture/models. Results in brackets are calculated considering detection errors as malware. They are considered benignware otherwise.}
\begin{tabular}{@{}cccc@{}}
\toprule
\multicolumn{2}{c}{\textit{Original Malware Samples}} & \multicolumn{2}{c}{\textit{Adversarial EXEmples}} \\ \toprule
\multicolumn{4}{c}{\textbf{No Calibration}} \\ \toprule
\textbf{Architecture} & \textbf{DR} & \textbf{Architecture} & \textbf{ADR} \\ \toprule
\pipeline            & 1.0 (1.0)   & \pipeline           &  0.86 (0.93)           \\
\pipeline No Dyn.    & 1.0 (1.0)   & \pipeline No Dyn.    & 0.76 (0.76) \\
\pipeline No Sign.     &   1.0 (1.0)          & \pipeline No Sign.     &  0.71 (0.89)         \\
Nebula                &   0.47 (0.82)          & Nebula                &  0.45 (0.79)           \\ \toprule
\multicolumn{4}{c}{\textbf{After Calibration}} \\ \toprule
\textbf{Architecture} & \textbf{DR} & \textbf{Architecture} & \textbf{ADR} \\ \toprule
\pipeline            & 1.0 (1.0)   & \pipeline           &  0.88 (0.95)\\
\pipeline No Dyn.    & 1.0 (1.0)   & \pipeline No Dyn.    & 0.81 (0.81)\\
\pipeline No Sign.     &   1.0 (1.0)          & \pipeline No Sign.     &  0.79 (0.93)\\
Nebula                &   0.55 (0.90)& Nebula                &  0.55 (0.89)\\ \toprule
\end{tabular}
\label{adversarial_results_sections}
\end{table}

While all untainted malware samples are detected by \pipeline, with the exception of Nebula alone due to the high number of emulation errors, results of transfer evaluation highlight an evasive trend against all models.
These suggest that GAMMA attack can evade the whole \pipeline, decreasing the Detection Rate by 14\%. 
This is achieved also thanks to the policy we established of flagging samples as benignware in case of errors of the last dynamic module. 
As we can see from the ADR if we count errors as malware, we derive that the 7\% of Adversarial EXEmples crashes the analysis, and the remaining 7\% bypasses \pipeline without pre-processing errors.
By looking at Nebula performances on these data, it is likely that roughly half of them are not detected at all, but also that some are, surprisingly, evading detection.
This can be noticed by looking again at the results scored by Nebula: when considering errors as malware, there is a small drop of performance (3\%) implying that those EXEmples do not crash the pipeline.
By inspecting the results, we discovered that (i) some malware samples obtain a different score when emulated multiple times inside a single Speakeasy session, and (ii) GAMMA changes the output score of Nebula.
We analyzed those reports, and we discovered that samples that invoke specific Windows APIs like:
\begin{itemize}
    \item \texttt{GetCurrentThreadId}, and \texttt{GetSystemTimeAsFileTime} get different return values when evaluated multiple times, thus changing the score. These are likely used to detect the presence of dynamic analysis, impacting the reports of emulation as well;
    \item \texttt{GetModuleFileNameW} retrieves a different file name. It is likely that Speakeasy rename the file to analyze with its hash. Thus, the output report of the untainted malware and its adversarial EXEmple counterpart are different. 
\end{itemize}
This analysis is not comprehensive, as it would need deeper reverse engineering of those samples, but, even if for few samples, we can conclude that also dynamic analysis is evaded by non-behavioral attacks, not strictly due to the effect of perturbations.
Lastly, 5 samples that originally caused errors during emulation are now analyzed by Speakeasy after the manipulation. 
This result needs further investigation, but they bring evidence that dynamic analysis based on emulation can be weak and discontinuous not only against adversarial transfer attacks, but also to multiple evaluations of the same samples.

If we remove dynamic analysis from \pipeline (\pipeline No Dyn. in \autoref{adversarial_results_sections}), we detect a drop of ADR by 10\%, implying that such a module was indeed able to stop some of the incoming attacks.
However, counter-intuitively, if we instead remove signatures (\pipeline No Sign.) and we keep dynamic analysis, we assist to a drop of ADR by 15\% w.r.t. \pipeline.
This means that, surprisingly, pattern-matching with YARA rules has a bigger impact on robustness than dynamic analysis itself.
This relevant contribution of YARA rules is due to how the section injection manipulation works:
while mixing together sections extracted from input data, it is likely that GAMMA creates some patterns that trigger YARA rules, contrary to the original PE.
We notice that 20 samples evade the YARA module before GAMMA manipulation, but they are detected once perturbed.
To better understand this result, we analyze a small subset of rules that are triggered by adversarial EXEmples, and we analyze them through expert domain knowledge.
Through this study, we isolate culprits of such triggers:
\begin{itemize}
    \item \texttt{CryptoLocker\_rule2}: it triggers on meta-data component of PE file.
    For instance, manifest.xml appears many times after section injection, instead of appearing only once (like in any regular PE), triggers the rule as a consequence;
    \item \texttt{AutoIT\_Compiled}: AutoIt is a rarely used scripting language for Windows\footnote{\url{https://www.autoitscript.com/site/}}, and it was found to be widely employed in malware crafting\footnote{\url{https://www.autoitscript.com/wiki/AutoIt_and_Malware}}. 
    This rule triggers whenever finds sections compiled by AutoIt. It happened that two sections were injected from a benign PE which has some AutoIt-compiled artifacts, and the rule was triggered;
    \item \texttt{SUSP\_NET\_NAME\_ConfuserEx}: Confuser is a well-known packer for .NET apps,\footnote{\url{https://github.com/yck1509/ConfuserEx}} used to avoid reverse engineering proprietary code.
    However, this tool is used also by malware developers to obfuscate malicious code, and this rule blocks sections which manipulated with Confuser. 
    In this case, some benign sections altered by Confuser were injected to craft the adversarial EXEmple;
    \item \texttt{Windows\_Trojan\_Njrat\_30f3c220}: it appears that some benign program contained in $\dstwo$, which is the dataset from which we extracted benignware for GAMMA, might not be actually legitimate. In this case, GAMMA injects a malicious payload that triggers the rule and prevents the adversarial EXEmple from being effective.
\end{itemize}

Thus, we can conclude that \pipeline is indeed robust against transfer adversarial EXEmples computed against its strongest component, but such robustness is achieved more by static signatures than dynamic analysis alone.
Moreover, we report another interesting effect of adversarial manipulation on signatures: some rules include a \texttt{filesize} condition to fire. If a PE is larger than a preset size, the rule is not fired. While this condition can be useful to avoid false alarms, it prevents some rules from being fired against adversarial EXEmples. 
Injected sections can increase the size of the manipulated malware samples beyond the limits set by certain rules, effectively allowing them to evade signature detection.

Calibration has a positive effect on robustness, increasing \pipeline's ADR by 2\%. As a side-effect, all the modules themselves exhibit an improved robustness, because the tuned thresholds contribute in blocking more adversarial EXEmples. 
In particular, the decrease of Nebula's threshold improves by 10\% its ADR, increasing its impact.

\mypar{GAMMA Padding Attack.}
From the same 1k malware considered for GAMMA section-injection, we compute 469 adversarial EXEmples with a padding manipulation, 81 more than GAMMA sections injection attack. 
This difference is because padding manipulation does not modify headers of malicious samples and \gbdt can be misled more easily. 
Indeed, EMBER features also consider the header of the PE, and padding manipulation can fool the computation of such features.
While \pipeline can detect all the original malicious samples, also when removing YARA or Nebula from its architecture, the GAMMA Padding attack decreases the Detection Rate by 13\% and by 11\% with stock and optimal thresholds respectively. 
Similarly to what we have observed for GAMMA section-injection, the calibration increases the robustness.
This is caused by the decrease of the decision threshold of \gbdt, requiring  attackers to create adversarial EXEmples with higher confidence. 
Regarding dynamic analysis, we notice the same discontinuous behaviors of Speakeasy emulation and Nebula prediction.
In particular, we notice that Nebula attributes different scores to the original malware and its adversarial EXEmple counterpart.
Also, two samples that originally crash emulation are correctly emulated after the attack.
Another pattern that presents is the bigger contribution of YARA to ADR w.r.t. Nebula, particularly when we test stock decision thresholds. 
On the contrary, with calibration, Nebula has a greater impact than YARA, because its optimal threshold is lower.
Contrary to sections injection attack, GAMMA padding can also succeed against YARA in two cases. 
We notice that two adversarial EXEmples are able to evade the first module, contrary to the original untainted malware.
More precisely, the rules which fire before the padding manipulation are: 
\begin{itemize}
    \item \texttt{SUSP\_XORed\_URL\_In\_EXE:} as the name tells, detects a XORed URL in an executable. The PE that triggers this rule downloads a trojan inside a system's directory, by connecting to a XOR-obfuscated URL; 
    \item \texttt{SUSP\_RANSOMWARE\_Indicator\_Jul20:}this rules detects the presence of specific strings like \texttt{DecryptFilesHere.txt} or \texttt{Decrypt.txt}, which are correalated with ransomware activities;\footnote{\url{https://securelist.com/lazarus-on-the-hunt-for-big-game/97757/}}
    \item \texttt{win\_gandcrab\_auto:} this rule aims detects specific instance of the \textit{win.gandcrab} malware, by looking for particular strings (automatically selected from the disassembly of memory dumps and unpacked files, using a tool named ``YARA-Signator'').\footnote{\url{https://malpedia.caad.fkie.fraunhofer.de/details/win.gandcrab}}
\end{itemize}
Another relevant pattern is that rules triggered by adversarial EXEmples are a subset of the ones triggered by original samples. 
Hence, it seems that padding prevents some rules from firing. With further investigation, we discover that also the three above-mentioned rules include a condition on the \texttt{filesize} as described in the analysis of GAMMA section-injection.
However, GAMMA padding manipulation can heavily affect PE size and consequently, they can evade signatures module as well. 
Hence, YARA rules can be beneficial for robustness, as seen with GAMMA sections-injection attacks, but easily bypassed by adding content if they present a limit on the file size.
\begin{table}[]
\caption{Adversarial Detection Rate (ADR) of diverse architectures/model under \textbf{GAMMA Padding} attacks. We compare not manipulated malware samples and the same samples after manipulation (Adversarial EXEmples), considering \textbf{stock} and \textbf{calibrated} architecture/models. Results in brackets are calculated considering detection errors as malware. They are considered benignware otherwise.}
\begin{tabular}{@{}cccc@{}}
\toprule
\multicolumn{2}{c}{\textit{Original Malware Samples}} & \multicolumn{2}{c}{\textit{Adversarial EXEmples}} \\ \toprule
\multicolumn{4}{c}{\textbf{No Calibration}} \\ \toprule
\textbf{Architecture} & \textbf{DR} & \textbf{Architecture} & \textbf{ADR} \\ \toprule
\pipeline            & 1.0& \pipeline           &  0.87 (0.95)\\
\pipeline No Dyn.    & 1.0& \pipeline No Dyn.    & 0.80 (0.80)\\
\pipeline No Sign.     &   1.0& \pipeline No Sign.     &  0.77 (0.94)\\
Nebula                &   0.45 (0.81)& Nebula                &  0.44 (0.80)\\ \toprule
\multicolumn{4}{c}{\textbf{After Calibration}} \\ \toprule
\textbf{Architecture} & \textbf{DR} & \textbf{Architecture} & \textbf{ADR} \\ \toprule
\pipeline            & 1.0& \pipeline           &  0.89 (0.95)\\
\pipeline No Dyn.    & 1.0& \pipeline No Dyn.    & 0.80 (0.80)\\
\pipeline No Sign.     &   1.0& \pipeline No Sign.     &  0.81 (0.94)\\
Nebula                &   0.52 (0.88)& Nebula                &  0.54 (0.89)\\ \toprule
\end{tabular}
\label{adversarial_results_padding}
\end{table}
\begin{tcolorbox}[colback=red!40, colframe = black, boxrule=0.5pt]
    \textbf{Take-home message 5:} \pipeline can withstand adversarial transfer attacks, but the effect of dynamic analysis is counter-intuitively marginal.
    Attacks can trigger YARA rules that were not predictive against the original untainted samples, since GAMMA introduces artifacts. 
    Also, some rules are disabled by the attack, due to constraints on filesize which are now violated by attacks.
\end{tcolorbox}
\subsection{Computation Times Comparison}
We now analyze the classification times of all the modules and models tested, also considering the calibrated \pipeline, and we report our findings in \autoref{tab:times} to answer \textbf{RQ6}.
We measure MDT for malware and benignware separately on both $\dsone$, and $\dstwo$, as explained in \autoref{sec:exp_results}, including the standard deviation of each measurement as well.
In general, we can see a difference between times computed for malware (first column) and benignware (second column).
This discrepancy is mainly attributed to the larger dimension of the benignware in terms of bytes and content included in programs, thus slowing down all the compared techniques.
As expected, signatures and static models are the fastest techniques, with \malconv being the best one, followed by YARA pattern-matching and \gbdt.
Intuitively, \malconv does not pre-process input samples, and it just uses their first 1MB to compute predictions without relying on feature extractors.
On the contrary, while pattern-matching is indeed fast, it is slowed down by the length of the file to analyze, requiring different amounts of computations for smaller and larger programs.
Lastly, \gbdt is two orders of magnitude slower than \malconv due to the heavy EMBER feature extraction phase.
Unsurprisingly, hybrid and dynamic analysis with Quo.Vadis and Nebula requires a non-negligible amount of seconds due to emulation.
On the contrary, even if built on top of many components, \pipeline rivals \gbdt when analysing malware.
Since computations are halted at first detection, \pipeline cuts the heavy feature-extraction phase when possible.
On the contrary, since benign samples must traverse all the pipeline, we report a slightly higher required time to analyze benignware, mostly due to the emulation step.
Thus, we can conclude that the overhead induced by \pipeline is negligible w.r.t. both static and dynamic analysis alone. 
However, on calibrated \pipeline, we highlight that, on average, samples require more time to be analyzed on average.
Since \malconv decision threshold increases, more malicious samples are sent to the next modules. At the same time, due to the reduction of false positives, more benignware do not trigger false alarms and they are analyzed by all the components of \pipeline.
Together, these two facts lead to an increase in the average analysis time, both for benign and malicious samples.
\begin{table}[]
\caption{Mean Detection Time (MDT) of each module or architecture testes, separately computed for malware and benignware subsets of the original datasets.}
\begin{tabular}{@{}ccc@{}}
\toprule
\textbf{Architecture / Module} & \textbf{MDT$_m$ (s)}         & \textbf{MDT$_g$ (s)}         \\ \toprule
Signatures                     &  $2.9\pm 6.8 \times 10^{-2}$ &  $6.5\pm 18 \times 10^{-2}$  \\
\malconv                        &  $8.9\pm 20 \times 10^{-3}$ &    $6.5\pm 3.1 \times 10^{-3}$  \\
GBDT                           &  $1.5\pm 4.0 \times 10^{-1}$    &   $2.2\pm 6.4 \times 10^{-1}$ \\
Nebula                         &  $2.4 \pm 16$        &      $3.0 \pm 14$     \\
Quo.Vadis                      &   $2.2 \pm 8.0$      &    $6.0 \pm 18$       \\ \toprule
\multicolumn{3}{c}{\textbf{No Calibration}} \\ \toprule
\pipeline No Dyn.               &    $1.1\pm 3.6 \times 10^{-1}$   & $3.4\pm 8.4\times 10^{-1}$   \\
\pipeline             &     $2.0\pm 11 \times 10^{-1}$  & $6.3\pm 19$    \\ \toprule
\multicolumn{3}{c}{\textbf{After Calibration}} \\ \toprule
\pipeline No Dyn.               &    $1.2\pm 3.4 \times 10^{-1}$& $3.9\pm 9.7\times 10^{-1}$\\
\pipeline             &     $2.3\pm 14 \times 10^{-1}$& $6.6\pm 19$\\ \toprule
\end{tabular}
\label{tab:times}
\end{table}
\begin{tcolorbox}[colback=red!40, colframe = black, boxrule=0.5pt]
    \textbf{Take-home message 6:} Classification times strongly depend on the size of files, and the presence of emulation.
    The architecture of \pipeline does not introduce noticeable overheads while keeping all types of analyzes in one place. However, calibration induces a slightly-higher analysis time, due to the change of which module detect which malware.
\end{tcolorbox}

\section{Related Work}
\label{related}
After the establishment of Machine Learning models in the malware analysis domain, academic research has concentrated on analyzing ML models for malware detection, introducing numerous architectures for Machine Learning-based malware detectors that utilize signatures, static features, dynamic features, or a combination of them, known as hybrid analysis~\cite{gibert2020rise, ye2017survey}.
Our work can be labeled as the latter, which complements the information retrieved from static analysis with behavioral aspects retrieved during dynamic analysis. 
While our work focuses on sequential analysis, which keeps each analysis separate by submitting samples to different classifiers (each trained on distinct features) in a step-by-step sequence, previous work considered mixing both static and dynamic features mainly in two ways.
Early fusion strategy merges all the features before training a single classifier. Instead, in late fusion strategy, separate classifiers for each family of features are trained and then the results are combined in a single score. 
We summarize in \autoref{related works} a pool of the most relevant works that studied how to combine features from different malware analyzes, by annotating each work with the applied hybrid analysis employed (early fusion, late fusion or sequential), the source of the selected features (from malware signatures, static or dynamic analysis) and we report the analysis presented to assess the quality of the described architectures. For this last column, we report keywords for brevity and readability.

As already mentioned in this work, with Quo.Vadis,~\cite{trizna2022quo} developed a deep learning model architecture that analyze separately static and dynamic features and then merges results with a single meta-model through late fusion.
\cite{yen2019integration} mix several features computed from both static and dynamic analysis to build early fusion models, late fusion models, and then a mixed fusion neural network. Also~\cite{ma2016using} use an ensemble of different classifiers to perform a hybrid analysis building training models with separate feature sets made coupling static and dynamic analysis telemetry.
\cite{shijo2015integrated} document a way to construct a feature vector leveraging data from static and dynamic analysis techniques and processed by a single ML model.
~\cite{han2019maldae} and ~\cite{han2019malinsight} tested different ML models in malware detection and classification tasks, merging static and dynamic features, and training models directly with the whole set of mixed features. The same approach is described by~\cite{kumar2019malware}, which also report an interesting analysis on virtualization times: they study how accuracy varies adjusting the run time of samples in a sandbox. All these works apply the so-called early fusion strategy, and experiments report an improvement in accuracy using both families of features.
~\cite{ngo2023fast} tackled the problem of feature fusion with Transfer Learning and Knowledge Distillation (KD): a large \textit{teacher model} was trained on aggregated static and dynamic features and a small \textit{student model} only on static features. Then the knowledge of the rich "behavior-aware" model is transferred to the faster small model to perform classification. Similarly to our work, this approach aims at reducing detection delays caused by dynamic analysis: indeed, after the training effort of the teacher model including behavioral features, this knowledge is only transferred to the student model, which computes only static features to perform classification. 

Different from most of the works mentioned before, we also focus on evaluating the robustness of \pipeline to adversarial EXEmples, analyzing the contribution of dynamic model and signatures on ADR; moreover, we present two different policies for impossible-to-analyze samples, while in literature they are discarded or they are not the object of interest for any further analysis. Moreover, as highlighted by \autoref{related works}, as far as we know, we are the first to analyze a sequential technique that chains together different kinds of analysis, considering error-handling, robustness, and calibration of those modules.

\begin{table*}[]
\caption{Overview of Related Works described in \autoref{related}. We describe the type of hybrid analysis methodology employed, the sources of features and malware characteristics used to construct the dataset. We also provide the keywords for the analyzes performed in the experimental evaluation to test the proposed architectures.}
\label{related works}
\resizebox{\linewidth}{!}{
\begin{tabular}{@{}c|ccc|ccc|cccc@{}}
 &
  \multicolumn{3}{c|}{\textbf{Methodology}} &
  \multicolumn{3}{c|}{\textbf{Malware Analysis Type}} &
  \multicolumn{4}{c}{\textbf{Discussed Analysis}} \\ \toprule
  \textbf{Work} &
  \textbf{Early Fusion} &
  \textbf{Late Fusion} &
  \textbf{Sequential} &
  \textbf{Signatures} &
  \textbf{Static} &
  \textbf{Dynamic} &
  \textbf{Keywords} &
  \textbf{} &
  \textbf{} &
  \textbf{} \\ \toprule
  \cite{trizna2022quo} &  & \checkmark &  &  & \checkmark & \checkmark & Performance, Robustness \\
 \cite{yen2019integration} & \checkmark & \checkmark &  &  & \checkmark & \checkmark & Performance  \\
 \cite{shijo2015integrated}& \checkmark &  &  &  & \checkmark & \checkmark & Performance   \\ 
 \cite{ma2016using} &  & \checkmark &  &  & \checkmark & \checkmark &  Performance \\ 
 \cite{han2019maldae}& \checkmark &  &  &  & \checkmark & \checkmark &  Performance  \\ 
 \cite{han2019malinsight} & \checkmark &  &  &  & \checkmark & \checkmark &  Performance  \\ 
 \cite{kumar2019malware} & \checkmark &  &  & \checkmark & \checkmark & \checkmark & \makecell{Performance, Detection Times, \\ Virtualization Times}  \\ 
 \cite{ngo2023fast} & \checkmark &  &  &  & \checkmark & \checkmark &  Performance, Detection Times  \\
 \pipeline &  &  & \checkmark & \checkmark & \checkmark & \checkmark &  \makecell{Performance, Error Handling, Robustness, \\ Model Calibration, Detection Times} \\ 
\bottomrule
\end{tabular}}
\end{table*}

\section{Limitations}
\label{sec:limitations}
We now evaluate the limitations of our work, by discussing whether to handle them or their irrelevance. 

\mypar{\pipeline Model Training.}
Our pipeline's architecture is composed of pre-trained objects on different datasets. The scope of our work is not to build a new model to optimize a common loss function, training the whole pipeline on the same dataset. Still, we want to highlight the performances of a sequence of ML models in detecting malware, adding signatures to static and dynamic analysis, similarly to industrial architectures, analysing the contributions of each module. Building a single model which can analyze different features without relying on feature fusion as Quo.Vadis~\citep{trizna2022quo} is a non-trivial problem out of the scope of this work.

\mypar{\pipeline Calibration.}
In this work, we leverage pre-trained models and we propose a tentative fine-calibration of decision thresholds. In doing so, we perform an exhaustive grid search with different values models' thresholds inside \pipeline. As mentioned above, we do not have a common loss function to minimize in an end-to-end learning process, and this does not let us perform a proper cross-validation or other fine-tuning techniques.

\mypar{Temporal Analysis.}
We do not perform a systematic temporal analysis, looking at performance decreases, and submitting new documented future data. In literature, we find strategies to overcome temporal bias avoiding the so-called \textit{concept drift} problem~\citep{pendlebury2019tesseract}, i.e. the obsolescence of ML models trained on past data at test-time with an unseen dataset caused by the decay of i.i.d (independent and identically distributed) assumption on data.
However, MalConv and GBDT are trained on samples up to 2018, while Nebula training data is collected in January 2022, and $\dstwo$ was collected in April 2022: this means that our evaluation can be still treated as realistic.
Also, $\dsone$ is strongly imbalanced in terms of malware families, as frequently happens in a real-world scenario. Also, we do not have precise timestamps of this collection and we do not use it to make conclusions on temporal analysis.

\mypar{Evaluation of Analysis Errors.}
In our analysis, we report that labeling errors as benign helps reduce FPR, as demonstrated by experimental work. However, in a real-world scenario this could be harmful since an error can bypass \pipeline's detection in its actual configuration. While in this work we propose a first basic policy of error handling, we discuss more fine-grained solutions in \autoref{sec:conclusions}.

\mypar{Dynamic Analysis on Virtualization.}
Nebula model is trained on reports generated through Speakeasy~\citep{speakeasy}, that is an emulation tool. We see limitations in PE emulation, especially in reliability, as assessed by our experimental work. 
Virtualizing a huge amount of PE as done for the Speakeasy dataset~\citep{trizna2022quo} needs time and a well-settled environment and it is out of our purpose, which is testing available ML models for dynamic analysis inside a sequential pipeline.

\mypar{End-to-End Adaptive Attacks.}
We perform transfer attacks targeting only GBDT, namely the best-performing one. Attacks on ML models trained on static features are well implemented and documented~\citep{demetrio2021functionality}. 
However, while the evasion of dynamic analysis is a wide and well-known problem~\citep{afianian2019malware}, currently no open source code has been released to re-evaluate previously-proposed techniques.
As far as we know, there is only one attack that has been proposed by~\citep{rosenberg2018generic}, but it is not replicable since no code has been provided.
Moreover, we do not customise GAMMA to avoid including any artifact detected by YARA rules. 
However, our goal for this paper is the empirical analysis of malware pipelines that resembles industry settings, without delving into adversarial robustness, as this would require a deeper investigation on its own, as well as developing novel or re-implementing attacks.

\section{Conclusions}
\label{sec:conclusions}
In this work, we are the first to study the behavior of sequential malware detection pipelines in terms of performance, resilience to errors, and robustness.
To do so, we develop \pipeline, a Windows malware detector pipeline, that sequentially puts together static and dynamic analysis, leveraging pattern-matching with YARA and state-of-the-art machine learning models for malware detection.
We build an architecture that mimics industrial solutions, testing its performances compared to hybrid approaches, proposing policies to handle analysis errors, and analysing its robustness against adversarial EXEmples. We highlight interesting findings remarked by our experimental work: \pipeline proves to outperform state-of-the-art models designed for static, dynamic, or hybrid analysis, keeping the highest TPR with low FPR. Moreover, it proves faster detection times of malicious PEs as compared to models leveraging dynamic analysis; in achieving this, our proposed architecture does not discard any input sample and propagates detection errors until the end of the pipeline, and labeling impossible-to-analyze samples as benign proves to be the best policy to provide less false alarms.
Moreover, we present a simple way to calibrate decision thresholds of a series of pre-trained models, performing an exhaustive grid search on a validation set. We state that the calibration effort can bring advantages in terms of TPR and FPR in most scenarios, minimally sacrificing fast detection times.
\pipeline exhibits robustness against adversarial transfer attacks, also surprisingly demonstrates that GAMMA section-injection is detected more by YARA rules than by dynamic analysis, whose integrity is mined by small changes in the sample structure that are reflected in the emulation report.
Overall, we report the small contribution of emulation-based dynamic analysis to accuracy and robustness, besides the computational burden that is required. This proves of the effectiveness of our choice to use that kind of analysis only when required and as the last resort. 
Lastly, we underline the benefits of calibration on robustness, highlighting how \pipeline blocks more adversarial EXEmples generated with section injection and padding thanks to the tuning of the thresholds of the pre-built \gbdt, MalConv, and Nebula.

\mypar{Future Work.} As mentioned in \autoref{sec:limitations}, we did not learn all the parameters of the components of \pipeline, but rather we leveraged pre-trained models.
However, we believe that it is possible to formalize an algorithm that wraps all modules, whose parameters can be learned within a single minimization by also keeping static and dynamic analyzes separated. 
In addition, we want to improve the calibration mechanism, including other parameters in the grid, or other metrics to compute on the validation set, such as the detection time. Or, after finding a process to perform end-to-end training, we can perform proper cross-validation or other fine-tuning techniques.
We would want to improve the empirical error handling technique we deployed, by considering two alternatives. 
The first one considers relying on end-to-end models like \malconv as they are error-proof due to the lacking of feature extraction process. 
Hence, we could design an ensemble of end-to-end learners to give a more accurate decision on impossible-to-analyze samples. 
Thus, if further components of the pipeline would get an error, the decision would be taken by this ensamble instead.
Connected to this solution, a second alternative is to implement conformal prediction~\citep{vovk2005algorithmic, shafer2008tutorial}.
Error-proof models can be tuned using a conformal prediction framework~\citep{angelopoulos2023conformal} to know when (at which probability threshold) such models are certain in labeling a sample as benign, thus reducing the numbers of later errors.
Also, another future investigation will take into account the difference between emulation and virtualization.
In particular, we will train the dynamic analysis module on virtualization reports collected from a training PE dataset, and we will investigate the difference in both performance and overhead.
As regards end-to-end attacks, we plan on developing novel adversarial attacks suited to evade end-to-end malware pipelines like \pipeline, thus trying to deceive both malware signatures and dynamic modules.

\section*{Acknowledgments}
Andrea Ponte acknowledges the support of Rina Consulting S.p.A. to his doctoral scholarship and research work.
This work was partially supported by projects SERICS (PE00000014) and FAIR (PE00000013) under the NRRP MUR program funded by the EU - NGEU.

\appendix

\bibliographystyle{cas-model2-names}

\bibliography{cas-refs}

\end{document}